\documentclass{ws-procs9x6}
\usepackage{subfigure}
\begin{document}

\title{Final Results of the GEp-III Experiment and the
  Status of the Proton Form Factors}

\author{A. J. R. Puckett$^*$}

\address{Los Alamos National Laboratory,\\
Los Alamos, NM 87544, U. S. A.\\
$^*$E-mail: puckett@jlab.org\\
www.lanl.gov}

\author{The GEp-III Collaboration}
\address{Thomas Jefferson National Accelerator Facility, \\
  Newport News, VA 23606}



\begin{abstract}
The recently published final results of experiment E04-108 in
Jefferson Lab Hall C extend the recoil polarization measurements of
the ratio of the proton electric and magnetic form factors to $Q^2 =
8.5$ GeV$^2$, an increase in $Q^2$ coverage of more than 50\%. A global fit of
$G_E^p$ and $G_M^p$ to selected data for electron-proton elastic
scattering cross sections and polarization observables 
is presented, illustrating the statistical impact of the new results. 
\end{abstract}


\bodymatter


\paragraph{}
The elastic electromagnetic form factors of the nucleon are among the
most fundamental quantities describing its structure. In the
one-photon-exchange or Born approximation, the
helicity-conserving Dirac form
factor $F_1(Q^2)$ and the helicity-flip Pauli form factor $F_2(Q^2)$
fully characterize the amplitude for elastic electron-nucleon
scattering at a spacelike momentum transfer $Q^2 = -q^2>0$. These form factors
are accessible experimentally through both cross section and
double-polarization measurements, and are often expressed in terms
of the equivalent linear combinations $G_E = F_1 - \tau F_2$
(electric) and $G_M = F_1 + F_2$ (magnetic), which are more readily
extracted from experimental observables. The Born differential cross section for
elastic $ep$ scattering is given by\cite{Rosenbluth1950}
\begin{eqnarray}
  \frac{d\sigma}{d\Omega_e} &=& \frac{\alpha^2}{Q^2}
  \left(\frac{E'_e}{E_e}\right)^2\cot^2 \frac{\theta_e}{2}\frac{G_E^2 +
    \frac{\tau}{\epsilon}G_M^2}{1+\tau} \nonumber \\
  &=& \frac{\sigma_{Mott}}{\epsilon(1+\tau)} \left[\epsilon G_E^2 +
    \tau G_M^2 \right] \label{RosenbluthFormula}
\end{eqnarray}
where $\alpha$ is the fine-structure constant of electromagnetism, $E_e$ and $E'_e$
are the lab incident and scattered electron energies, $\theta_e$ is
the lab electron scattering angle, $\tau = Q^2 / 4M^2$, $M$ is the
nucleon mass, and $\epsilon = \left[1+2(1+\tau)\tan^2 \frac{\theta_e}{2} \right]^{-1}$ corresponds to the longitudinal
polarization of the virtual photon in the single-photon-exchange
picture.

The polarization of protons scattered elastically by longitudinally
polarized electrons has nonvanishing longitudinal and
transverse components relative to the momentum transfer in the
scattering plane given by \cite{AkhiezerRekalo2,ArnoldCarlsonGross}
\begin{eqnarray}
  I_0 P_l &=& \sqrt{\tau (1+\tau)} \tan^2 \frac{\theta_e}{2}
  \frac{E_e+E'_e}{M} G_M^2, \\
  I_0 P_t &=& -2 \sqrt{\tau(1+\tau)} \tan \frac{\theta_e}{2} G_E G_M,
\end{eqnarray}
where $I_0 = G_E^2 + \frac{\tau}{\epsilon} G_M^2$. The
component $P_n$ normal to the scattering plane is exactly zero in the Born approximation. The ratio
$P_t/P_l$ is directly proportional to $G_E/G_M$:
\begin{eqnarray}
  R = \mu_p \frac{G_E^p}{G_M^p} &=& -\mu_p \frac{P_t}{P_l}
  \frac{E_e+E'_e}{2M} \tan \frac{\theta_e}{2} \label{Rformula}
\end{eqnarray}
The ratio \eqref{Rformula} provides enhanced sensitivity to $G_E$ at
large $Q^2$, where $G_M$ dominates the cross section (see equation \eqref{RosenbluthFormula}). Precise recoil polarization experiments
\cite{Jones00,Punjabi05,Gayou02} first established that $R$
decreases rapidly with $Q^2 \ge 1$ GeV$^2$, in strong disagreement with
cross section measurements. The extraction of a small $G_E^2$ term compared to
a dominant $\frac{\tau}{\epsilon} G_M^2$ term from L/T
separation measurements becomes highly sensitive to incompletely understood
higher-order QED effects at large $Q^2$, including two-photon
exchange (TPEX)\cite{TPEXreview}. The enhanced sensitivity of the
recoil polarization method to $G_E$ and the typically smaller relative
importance of radiative corrections and TPEX to the ratio \eqref{Rformula} have
led to a general consensus that the 
polarization data most reliably determine $G_E^p$ at large $Q^2$. 

Extending the accurate measurements of the proton and neutron electric and
magnetic form factors to the highest accessible $Q^2$ provides
crucial experimental input to understanding the transition between the
non-perturbative and perturbative regimes of QCD. In addition to
constraining QCD-inspired phenomenological models, precise form factor
data provide important
model-independent information about the structure of the nucleon. The
recently published results of experiment E04-108\cite{Puckett2010} in Jefferson Lab's
Hall C extended the recoil polarization data for $R$ to $Q^2 = 8.5$
GeV$^2$. In this paper we illustrate the statistical impact of the
new data in a global fit to elastic electron-proton
scattering data, including both cross section and polarization measurements, using the Kelly parametrization\cite{KellyParametrization} of $G_E^p$ and
$G_M^p$.

The Kelly parametrization of the nucleon form
factors\cite{KellyParametrization} is given by the ratio of a
polynomial of order $n$ to a polynomial of order $n+2$ in $\tau =
Q^2/4M^2$:
\begin{eqnarray}
  G(Q^2) &=& \frac{\sum_{k=0}^{n} a_k \tau^k}{1+\sum_{k=1}^{n+2}b_k
    \tau^k}
  \label{KellyFitFormula}
\end{eqnarray}
This parametrization satisfies $G(Q^2) \propto Q^{-4}$ as $Q^2
\rightarrow \infty$ and gives the appropriate static ($Q^2 \rightarrow
0$) limit if one fixes $a_0= 1$ for $G_E^p$ and $G_M^p/\mu_p$. This 
parametrization is chosen to have the large-$Q^2$
behavior expected from dimensional scaling laws in perturbative
QCD\cite{BrodskyFarrar1975,BrodskyLepage1979}. The
original 2004 analysis by Kelly using this parametrization found that
a four-parameter fit with $n=1$ could describe the data for $G_E^p$
and $G_M^{p,n}$ with a reasonable $\chi^2$. For simplicity's sake, the
results presented here use the same choice.

The published elastic $ep$ scattering data chosen for the fit include 421
differential cross section measurements spanning $0.005 \mbox{ GeV}^2
\le Q^2 \le 31.2 \mbox{ GeV}^2$ and 53 polarization measurements
spanning $0.162 \mbox{ GeV}^2 \le Q^2 \le 8.49 \mbox{ GeV}^2$. The
differential cross section data were taken from Refs.
\refcite{Janssens66,Berger71,Price71,Bartel73,Kirk1973,Bork74,Bork75,Simon80,Sill93,Andi94,Christy04,Qattan05},
while the polarization data were taken from
Refs. \refcite{Alguard76,Milbrath98,Pospischil01,Punjabi05,Gayou01,Gayou02,Crawford07,Jones06,Ron07,Puckett2010}. While
by no means comprehensive, the chosen data are sufficiently representative of the
precision and $Q^2$ coverage of all elastic $ep$ scattering
data. All data included in this
analysis were obtained from hydrogen targets.

The best fit parameters for $G_E^p$ and $G_M^p$ were obtained in a
simultaneous global analysis of cross
section and polarization data by minimizing the $\chi^2$ function
defined as 
\begin{eqnarray}
  \chi^2 &=& \sum_{i=1}^{N_\sigma} \left[\frac{\sigma_{R,data}^{(i)} -
    \left(\epsilon G_E^2 + \tau G_M^2 \right)^{(i)}}{\Delta \sigma_{R,data}^{(i)}
  }\right]^2 + \sum_{i=1}^{N_{pol.}}
\left[\frac{R_{p,data}^{(i)}-R_p^{(i)}}{\Delta
    R_{p,data}^{(i)}}\right]^2 \label{chi2func},
\end{eqnarray}
where $R_p = \mu_p G_E^p/G_M^p$, $G_E^p$ and $G_M^p/\mu_p$ are given by the
formula \eqref{KellyFitFormula} with $n=1$ and $a_0 = 1$, $a_1^{E,M}$
and $b_{1,2,3}^{E,M}$ are eight adjustable parameters to be determined, 
$\sigma_{R,data} = \epsilon(1+\tau) \sigma_{data} / \sigma_{Mott}$
(see equation \eqref{RosenbluthFormula}), and $\Delta \sigma_{R,data}$ and $\Delta R_{p,data}$ are
the uncertainties in the experimental data. For the cross section
data, statistical uncertainties and overall normalization
uncertainties, when quoted with the published results, were included in $\Delta \sigma_{R,data}$. For the
polarization data, only statistical uncertainties were included. 

The inconsistency between the Rosenbluth and polarization
data for $Q^2 \ge 1$ GeV$^2$ was handled using an iterative procedure. Three iterations of the fit
were performed. For all cross section data with $Q^2 \ge 1$ GeV$^2$,
the value of $G_E^p$ in equation \eqref{chi2func} was replaced by a fixed value of $G_E^p$
calculated from the results of the previous fit, or, in the case of
the first iteration, using the results of Kelly's 2004
analysis\cite{KellyParametrization}. This replacement has the effect
that for $Q^2 \ge 1$ GeV$^2$, $G_E^p$ is entirely determined by
polarization data, while $G_M^p$ is determined by cross
section data, with $G_E^p$ fixed by the polarization data for
$R_p$. For $Q^2 < 1$ GeV$^2$, cross section and polarization data
are treated on an equal footing for both $G_E^p$ and $G_M^p$. The
numerical minimization was carried out using the TMinuit class within
the ROOT libraries\cite{ROOTlibraries}. Typically, no significant
improvement of the fit was found after two iterations of the 
starting function for $G_E^p$. 

This fit procedure was performed for two data sets
identical in every respect except that the new JLab Hall C data for
$R_p$ were excluded from the first data set and included in the second
set. 
\begin{table}
  \tbl{\label{tablefitresults} Fit results for $G_E^p$ and
   $G_M^p/\mu_p$, with (``Old'') and without (``New'') the data of
   Ref. \refcite{Puckett2010}. $G\infty/G_D = 16M_p^4/\Lambda^2
   \left(a_1/b_3\right)$ is the asymptotic value of $G_E^p/G_D$ and
   $G_M^p/(\mu_pG_D)$. The ``standard'' dipole form factor is defined as $G_D =
   \left(1+Q^2/\Lambda^2\right)^{-2}$, with $\Lambda^2 =
 0.71$ GeV$^2$.}
  {  \begin{tabular}{|ccccc|}
      \hline & Old $G_E^p$ & Old $G_M^p$ & New $G_E^p$ & New
      $G_M^p$ \\ \hline
      $a_1$ & -.390 $\pm$ .081 & .075 $\pm$ .023 & -.299 $\pm$ .072& .081 $\pm$ .023 \\ 
      $b_1$ & 11.01 $\pm$ .10 & 11.14 $\pm$ .08 & 11.11 $\pm$ .09 & 11.15 $\pm$ .08 \\
      $b_2$ & 13.57 $\pm$ .84 & 18.44 $\pm$ .16 & 14.11 $\pm$ .83 & 18.45 $\pm$ .16 \\
      $b_3$ & 11.4 $\pm$ 4.5 & 5.12 $\pm$ .65 & 15.7 $\pm$ 4.2 & 5.31 $\pm$ .66 \\ 
      $G_\infty/G_D$ & -.84 $\pm$ .49 & .360 $\pm$ .064 & -.47 $\pm$ .23 & .376 $\pm$ .061 \\ 
      $Q^2(G_E^p = 0)$ & 9.0 & & 11.8 & \\ \hline
    \end{tabular}}
\end{table}
Table \ref{tablefitresults} shows the results of the fit with and
without the new Jefferson Lab results. The parameter errors quoted are
the standard errors calculated from the diagonal elements of the
covariance matrix. Significant correlations among the parameters for
each form factor were found. These correlations imply that the
uncertainties in the asymptotic form factor values and the
uncertainties in $G_E^p$ and $G_M^p$ at any given $Q^2$ cannot be
calculated directly from the parameter errors, but instead require the
full covariance matrix. On the other hand, the correlation
coefficients between
the parameters describing $G_E^p$ and those describing $G_M^p$ were
generally small, at the level of a few percent. Table \ref{diracpaulitable} shows the asymptotic values of $Q^6F_2^p$
and $Q^4 F_1^p$ for the fits with and without the new high $Q^2$
data. 

\begin{table}
  \tbl{\label{diracpaulitable} Asymptotic values of $Q^4 F_1$ and
    $Q^6 F_2$, with and without Ref. \refcite{Puckett2010}.}
   { \begin{tabular}{|cccc|}
      \hline Old $Q^6 F^p_{2,\infty}$ & New $Q^6 F^p_{2,\infty}$ & Old $Q^4F^p_{1,\infty}$ & New $Q^4F^p_{1,\infty}$ \\ \hline
      3.28 $\pm$ .93 & 2.69 $\pm$ .51 & .507 $\pm$ .090 & .529 $\pm$ .086 \\ \hline 
    \end{tabular}}
\end{table}

Table \ref{chi2table} illustrates the quality of the fit in terms of
$\chi^2$. Although the overall $\chi^2$ of the fit $\chi^2/n.d.f. =
828.4/466 = 1.78$ might be regarded as relatively poor, the extent to
which the reduced $\chi^2$ exceeds one per datum reflects the inconsistency
between the Rosenbluth and polarization methods, as shown in table
\ref{chi2table}. The polarization data and the
cross section data for $Q^2 \le 1$ GeV$^2$ are consistent
with the fit results for $G_E^p$ and $G_M^p$. Relative to the number
of data points, the cross section data for $1 \le Q^2 \le 10$ GeV$^2$
contribute disproportionately to $\chi^2$, since the
$\epsilon$-dependence of the measured cross sections disagrees with
the slope predicted by the polarization data.
\begin{table}
   \tbl{\label{chi2table} $\chi^2$ of the final global fit with new
    data included.}
  {  \begin{tabular}{|ccc|}
      \hline Observable & $N_{data}$ & $\chi^2$ \\ \hline 
      $\sigma_R$, $Q^2 \le 1$ GeV$^2$ & 269 & 293.8 \\ 
      $\sigma_R$, $1 < Q^2 \le 10$ GeV$^2$ & 142 & 483.3 \\
      $\sigma_R$, $Q^2 > 10$ GeV$^2$ & 10 & 3.96 \\
      $R_p^{pol.}$, all $Q^2$ & 53 & 47.4   \\ \hline 
      All data & 474 & 828.4  \\
      \hline   
    \end{tabular}
 }
\end{table}

\begin{figure}
  \begin{center}
    \subfigure[$G_E^p/G_D$\label{gepfiga}]{\includegraphics[width=.49\textwidth]{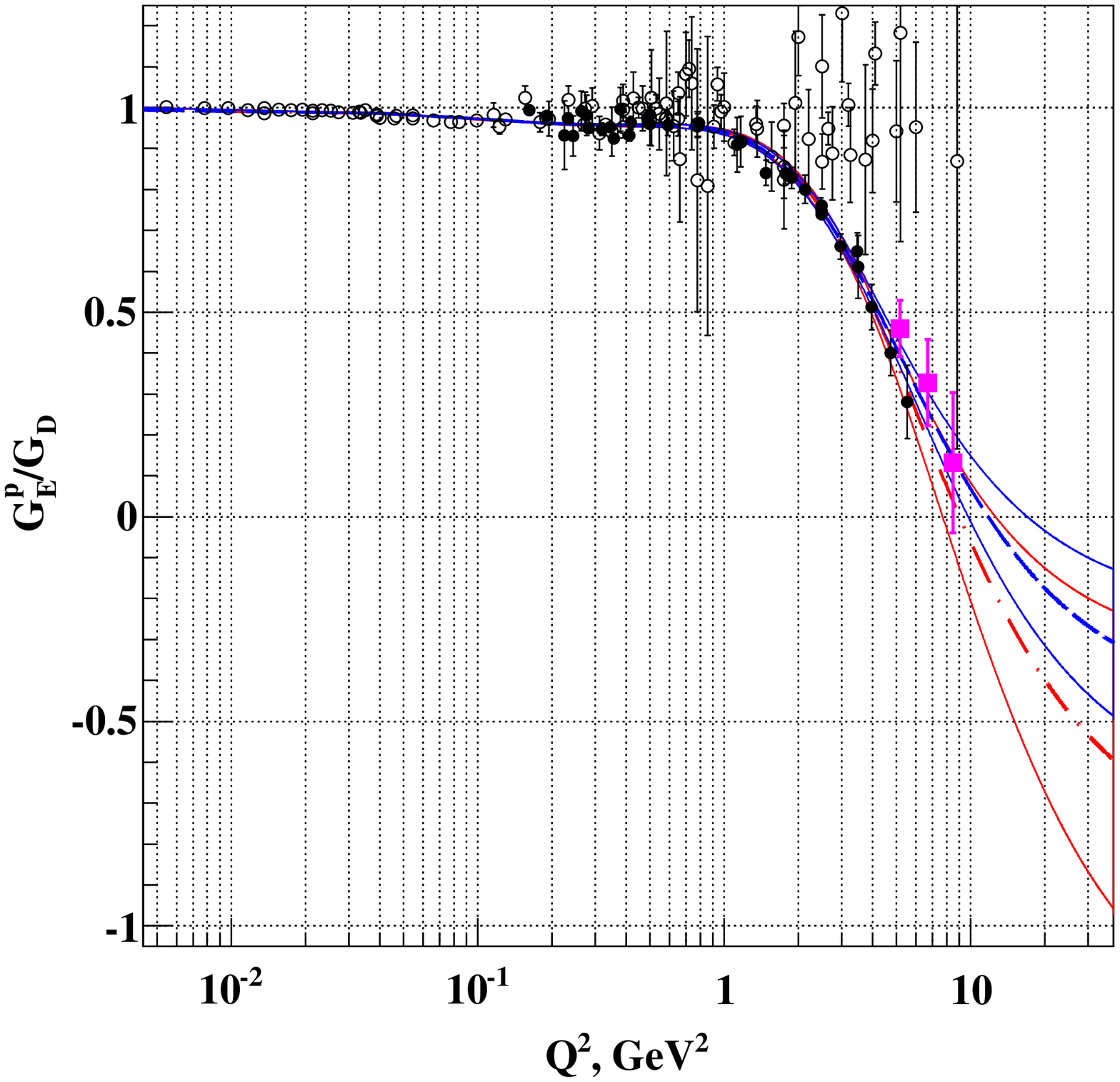}}
    \subfigure[$Q^6 F_2^p$\label{gepfigb}]{\includegraphics[width=.49\textwidth]{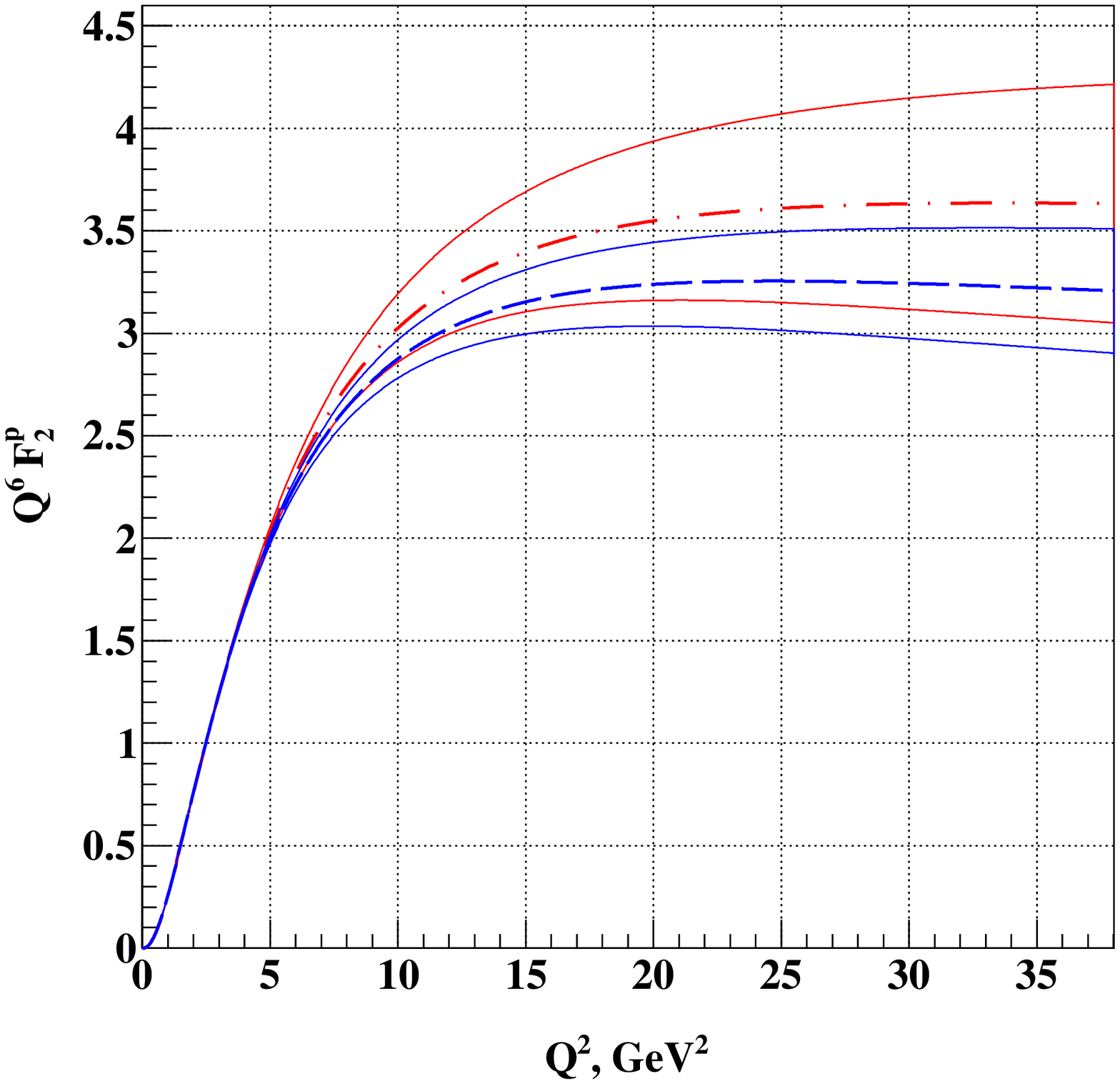}}
  \end{center}
  \caption{\label{gepfig} Fit results for $G_E^p/G_D$(a), and $Q^6 F_2^p$(b),
    before (red dot-dashed) and after (blue dashed) including the
    results of Ref. \refcite{Puckett2010}, with $1\sigma$ uncertainty
    bands. Published values of $G_E^p$
    extracted from cross section (open circles) and
    polarization (filled circles) measurements. Results of
    Ref. \refcite{Puckett2010} (magenta squares). }
\end{figure}

\begin{figure}
  \begin{center}
    \subfigure[\label{gmpfiga}$G_M^p/(\mu_pG_D)$]{\includegraphics[width=.49\textwidth]{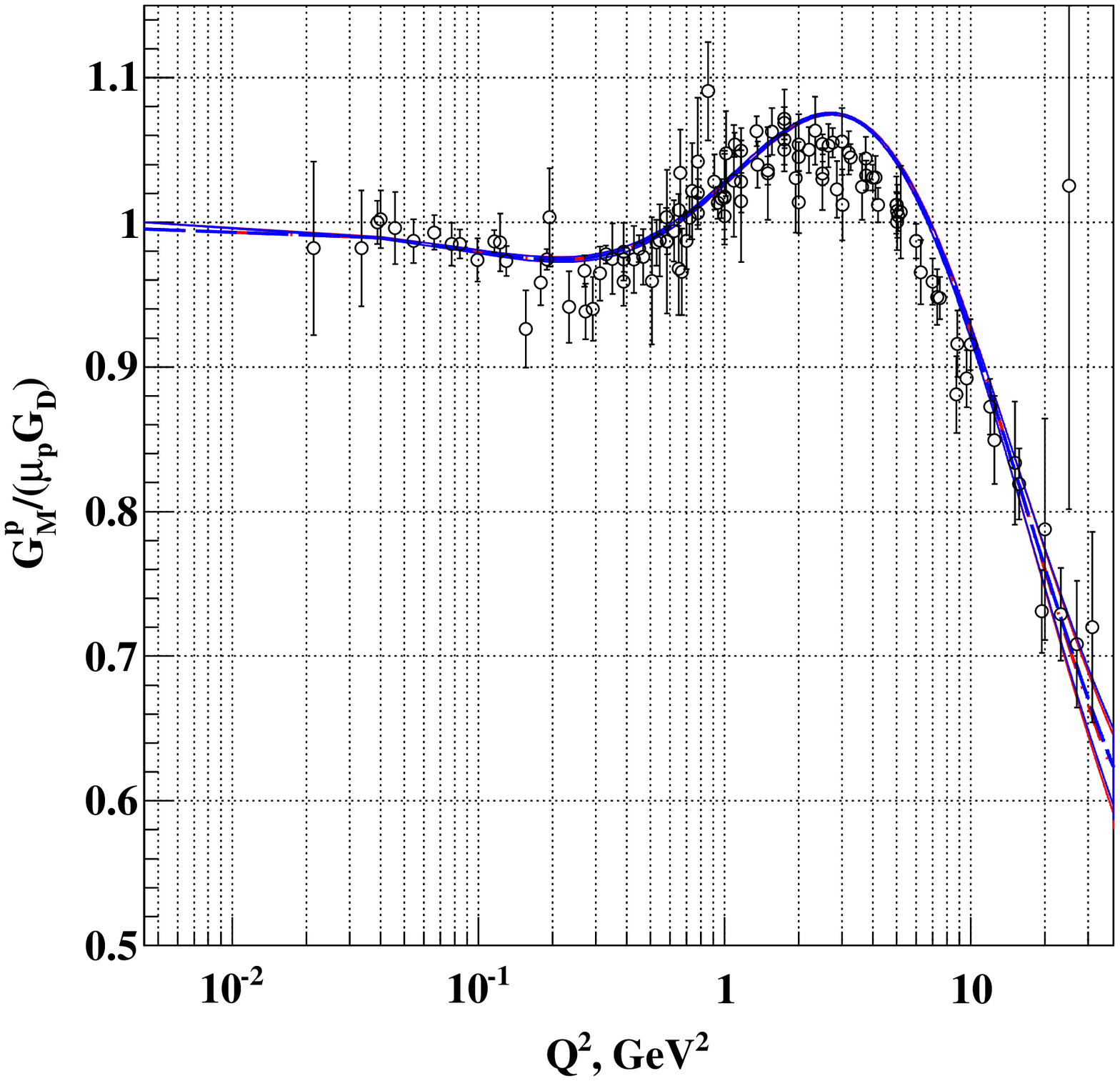}}
    \subfigure[\label{gmpfigb}$Q^4 F_1^p$]{\includegraphics[width=.49\textwidth]{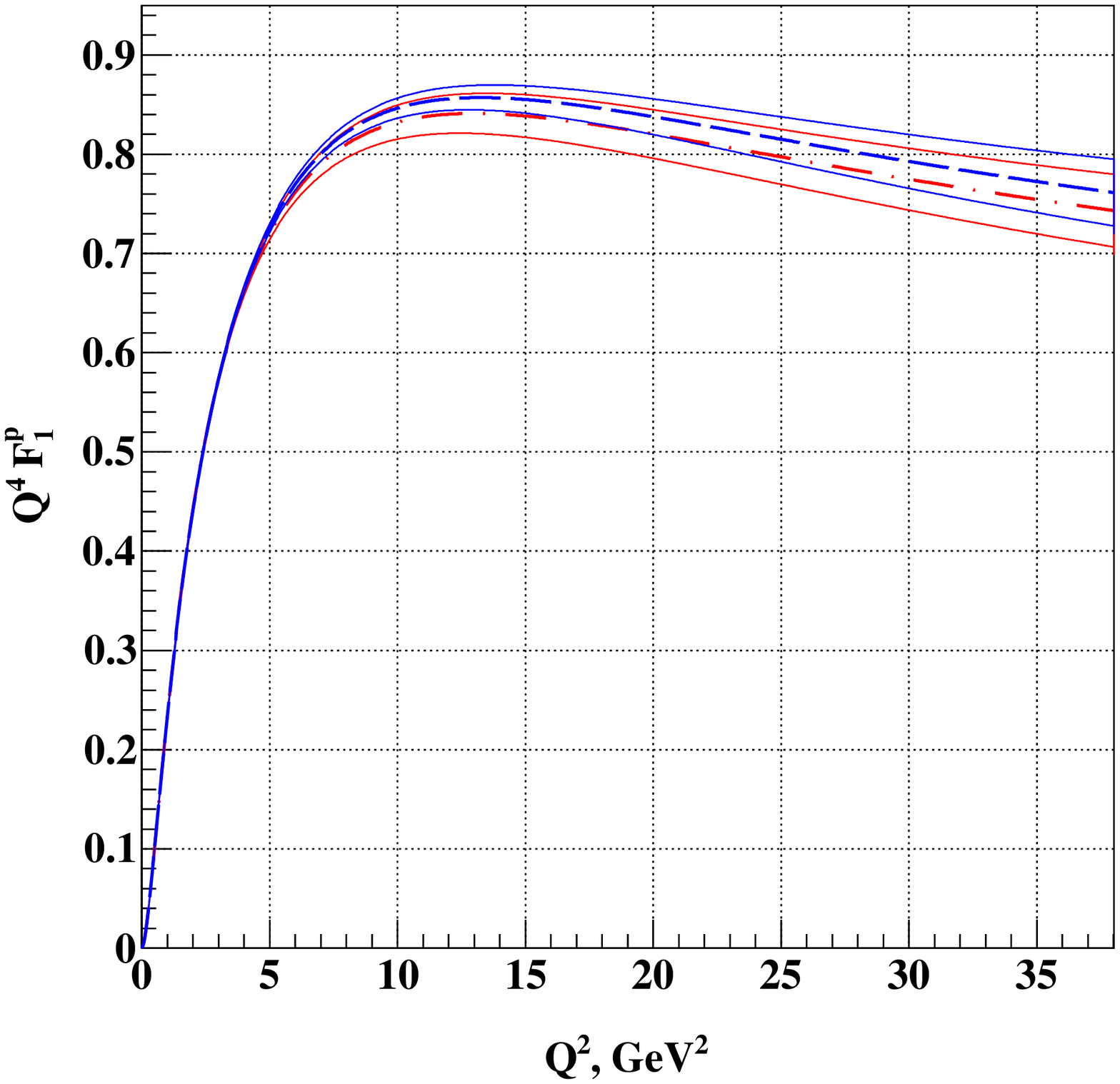}}
  \end{center}
  \caption{\label{gmpfig} Fit results for $G_M^p/(\mu_pG_D)$(a) and $Q^4 F_1^p$(b),
    before (red dot-dashed) and after (blue dashed) including the
    results of Ref. \refcite{Puckett2010}. Published values of $G_M^p$
    extracted from cross section (open circles) measurements.}
\end{figure}

Figures \ref{gepfig} and \ref{gmpfig} show the fit results together with
the experimental data. In figure \ref{gepfiga}, polarization data for
$R_p$ were converted to $G_E^p$ data using the fit result for
$G_M^p$ and appropriate error propagation. The best fit curves are
shown with standard $1\sigma$ uncertainty
bands calculated from the full covariance matrix. As shown in figure
\ref{gepfiga}, the new $G_E^p$ data\cite{Puckett2010} favor a
slowing rate of decrease of $G_E^p$ with $Q^2$ compared to previous
data, and shrink the uncertainty in $G_E^p$ for $Q^2 \ge 5$ GeV$^2$ by
roughly a factor of two. The improved constraint on $F_2^p$ in the
high-$Q^2$ region shown in figure \ref{gepfigb} is similarly dramatic. 

\begin{figure}
  \begin{center}
    \includegraphics[width=.49\textwidth]{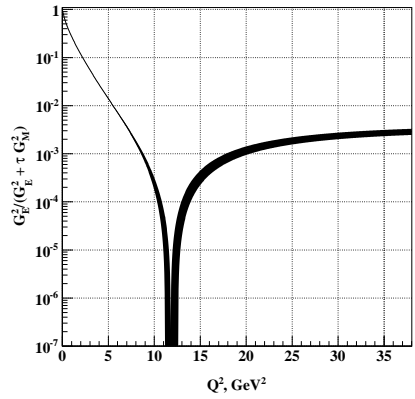}
  \end{center}
  \caption{\label{gefracfig}$G_E^2 / (G_E^2 + \tau G_M^2)$, calculated
    from the results of the global fit, represents
    the maximum fraction of the reduced cross section carried by $G_E$.}
\end{figure}

Figure \ref{gmpfig} shows the results for $G_M^p/\mu_p G_D$ and $Q^4
F_1^p$. The interesting feature of the $G_M^p$ results is that the fit
is systematically higher than the published data in the region of the
discrepancy. Since the cross section measurements either extracted or assumed $R_p \approx 1$, the
contribution of $G_E^p$ to $\sigma_R$ was overestimated, lowering the
extracted values of $G_M^p$. This result for $G_M^p$ is quantitatively
and qualitatively similar to the result of a recent global analysis including TPEX
corrections to the cross section data\cite{AMT2007}. In contrast
to this analysis, which neglects TPEX completely, the authors of Ref. [\refcite{AMT2007}] used the Kelly parametrization with $n=3$ and
achieved a substantially better $\chi^2$, since the applied TPEX
corrections largely reconciled the discrepancy. Although the new $R_p$ data barely affect $G_M^p$ and its uncertainty,
the effect on $F_1^p$ is more significant, an entirely expected result
given the dramatic improvement in $F_2^p$ and the definition $G_M =
F_1 + F_2$. For yet larger $Q^2$, the fit result for $G_M^p$ again
agrees with the published data, reflecting the dominance of $\sigma_R$
by $G_M^p$, shown in figure \ref{gefracfig}.

In conclusion, new high $Q^2$ measurements of $R = \mu_p G_E^p/G_M^p$
in Jefferson Lab's Hall C have significantly extended the range of
$Q^2$ for which $G_E^p$ is accurately determined. The impact of these new data
on the world database of proton form factors was studied in the
context of an empirical parametrization with physically reasonable assumptions for
low and high $Q^2$ asymptotic behavior; i.e., $G_E^p(0) = 1$ and
$G_M^p(0) = \mu_p$, and $F_{1(2)}^p \propto Q^{-4(6)}$ as $Q^2
\rightarrow \infty$.  In this context, the new data
for $R_p$ reduce the uncertainty in $G_E^p$ by roughly a factor of two
in the measured $Q^2$ region and in the extrapolation to higher
$Q^2$. Looking to the future, planned recoil polarization measurements
of $R_p$ to $Q^2 \approx 15$ GeV$^2$ and precision elastic $ep$
differential cross section measurements to $Q^2 \approx 18$ GeV$^2$ using JLab's upgraded 11 GeV
electron beam will complete the experimental knowledge of proton electromagnetic
form factors in
the spacelike region attainable with present day accelerators. 

\bibliographystyle{ws-procs9x6}
\bibliography{main}

\end{document}